\documentclass[prb,twocolumn,showpacs,preprintnumbers,amsmath,amssymb]{revtex4}

\usepackage{graphicx}
\usepackage{dcolumn}
\usepackage{bm}

\DeclareMathOperator{\coul}{coul}
\DeclareMathOperator{\erf}{erf}

\DeclareMathOperator{\Si}{Si}

\renewcommand{\Im}{\ensuremath{\text{Im\,}}}

\newcommand{\w}{\ensuremath{\omega}}

\renewcommand{\O}[1]{\ensuremath{\mathnormal{O}(#1)}}
\renewcommand{\b}[1]{\ensuremath{\mathbf{#1}}}
\renewcommand{\O}[1]{\ensuremath{\mathnormal{O}(#1)}}

\begin{document}

\title{Simple model of the static exchange-correlation kernel of a uniform electron gas with long-range electron-electron interaction}

\author{Julien Toulouse}
\email{toulouse@lct.jussieu.fr}
\affiliation{
Laboratoire de Chimie Th\'eorique, Universit\'e Pierre et Marie Curie, Paris, France.
}

\date{\today}
             
\begin{abstract}
A simple approximate expression in real and reciprocal spaces is given for the static exchange-correlation kernel of a uniform electron gas interacting with the long-range part only of the Coulomb interaction. This expression interpolates between the exact asymptotic behaviors of this kernel at small and large wave vectors which in turn requires, among other thing, information from the momentum distribution of the uniform electron gas with the same interaction that have been calculated in the $G_0 W_0$ approximation. This exchange-correlation kernel as well as its complement analogue associated to the short-range part of the Coulomb interaction are more local than the Coulombic exchange-correlation kernel and constitute potential ingredients in approximations for recent adiabatic connection fluctuation-dissipation and/or density functional theory approaches of the electronic correlation problem based on a separate treatment of long-range and short-range interaction effects.
\end{abstract}


\maketitle

\section{Introduction}

Density functional theory (DFT)~\cite{HohKoh-PR-64} applied within the Kohn-Sham (KS)~\cite{KohSha-PR-65} scheme is nowadays a widely-used method for electronic calculations in condensed-matter physics and quantum chemistry. The design of better approximations to the exchange-correlation energy functional, the central quantity of the theory, constitutes an important topic of research.

Following early ideas~\cite{NozPin-PR-58,KohHan-JJJ-XX,StoSav-INC-85}, the possibility of improving the description of the exchange and/or correlation energy functionals by treating separately its long-range and short-range components has recently gained growing interest~\cite{SavFla-IJQC-95,Sav-INC-96a,Sav-INC-96,LeiStoWerSav-CPL-97,PolSavLeiSto-JCP-02,PolColLeiStoWerSav-IJQC-03,SavColPol-IJQC-03,TouColSav-PRA-04,TouColSav-JCP-05,KohMeiMak-PRL-98,IikTsuYanHir-JCP-01,KamTsuHir-JCP-02,TawTsuYanYanHir-JCP-04,YanTewHan-CPL-04,HeyScuErn-JCP-03,HeyScu-JCP-04,BaeNeu-JJJ-XX}. In this approach, the Coulomb electron-electron interaction is decomposed as
\begin{equation}
\frac{1}{r} = v^{\mu}(r) + \bar{v}^{\mu}(r),
\label{}
\end{equation}
where $v^{\mu}(r)$ is a long-range interaction, $\bar{v}^{\mu}(r)$ is the complement short-range interaction and $\mu$ is a parameter controlling the range of the separation. For example, a convenient long-range interaction which has been often used is the so-called erf interaction~\cite{GilAdaPop-MP-96,Sav-INC-96,LeiStoWerSav-CPL-97,PolSavLeiSto-JCP-02,PolColLeiStoWerSav-IJQC-03,SavColPol-IJQC-03,TouColSav-PRA-04,TouColSav-JCP-05,IikTsuYanHir-JCP-01,KamTsuHir-JCP-02,TawTsuYanYanHir-JCP-04,YanTewHan-CPL-04,HeyScuErn-JCP-03,HeyScu-JCP-04}
\begin{equation}
v_{\erf}^{\mu}(r) = \frac{\erf(\mu r)}{r},
\label{}
\end{equation}
which vanishes for $\mu=0$ and reduces to the Coulomb interaction for $\mu \to \infty$. The Coulombic exchange-correlation energy functional $E_{xc,\coul}[n]$ can be in turn decomposed as~\cite{SavFla-IJQC-95,Sav-INC-96a,Sav-INC-96,LeiStoWerSav-CPL-97,PolSavLeiSto-JCP-02,SavColPol-IJQC-03,TouColSav-PRA-04,TouColSav-JCP-05}
\begin{equation}
E_{xc,\coul}[n] = E_{xc,\erf}^{\mu}[n] + \bar{E}_{xc,\erf}^{\mu}[n],
\label{Exc}
\end{equation}
where $E_{xc,\erf}^{\mu}[n]$ is the long-range component of exchange-correlation energy, associated to the interaction $v^{\mu}_{\erf}(r)$, and $\bar{E}_{xc,\erf}^{\mu}[n]$ is the complement short-range part. 

The long-range exchange-correlation energy, $E_{xc,\erf}^{\mu}$, can be efficiently calculated by configuration interaction (CI)~\cite{LeiStoWerSav-CPL-97,PolSavLeiSto-JCP-02} or multi-configurational self-consistent-field (MCSCF)~\cite{PedJen-JJJ-XX} methods enabling to handle near degeneracy in atoms and molecules, and by second-order perturbation theory~\cite{GerAngSavTou-JJJ-XX} or adiabatic connection fluctuation-dissipation (ACFD)~\cite{KohMeiMak-PRL-98,AngGerSavTou-JJJ-XX} approaches for describing van der Waals interactions. The short-range exchange-correlation energy $\bar{E}_{xc,\erf}^{\mu}$ can be well approximated by simple (semi)local density functional approximations~\cite{TouColSav-PRA-04,TouColSav-JCP-05} based a uniform electron gas model with a modified electron-electron interaction~\cite{Sav-INC-96,TouSavFla-IJQC-04} (see also Ref.~\onlinecite{ZecGorMorBac-PRB-04}).

Knowledge of the properties of the uniform electron gas is always useful to gain more insight into the physics of electronic correlation of inhomogeneous systems and ultimately to construct better approximations. In particular, in standard KS-DFT, the static exchange-correlation kernel of the Coulombic uniform electron gas, $f_{xc,\coul}$, is of interest and several accurate parametrizations are available (see, e.g., Refs.~\onlinecite{RicAsh-PRB-94,MorCepSen-PRL-95,CorDelOniPal-PRB-98}). Indeed, $f_{xc,\coul}$ has been recognized as a potential ingredient for density functional approximations (see, e.g., Refs.~\onlinecite{HohKoh-PR-64,KohSha-PR-65,GunJonLun-PRB-79,MorSen-PRB-91,Eng-PRA-95,LikMorSen-PRB-97,PalOniDelCorRei-PRB-99,OlePalOniDel-PRB-99}). $f_{xc,\coul}$ can also be used in principle in ACFD approaches (see, e.g., Ref.~\onlinecite{DobWan-PRB-00} for an example using a simplified kernel) or time-dependent density functional theory (TDDFT) for density-density response calculations, although in these cases clear strategies to use $f_{xc,\coul}$ are still under investigations. As a matter of fact, the high non-locality of $f_{xc,\coul}$ questions its transferability to inhomogeneous systems and limits its use.

In the framework of a long-range/short-range energy decomposition in DFT, knowledge of the static exchange-correlation kernel $f_{xc,\erf}^{\mu}$ of a uniform electron gas with the long-range erf interaction $v^{\mu}_{\erf}(r)$, as well as its complement analogue $\bar{f}_{xc,\erf}^{\mu} = f_{xc,\coul} - f_{xc,\erf}^{\mu}$ associated to the short-range part of the Coulomb interaction, is also of interest. In particular, $f_{xc,\erf}^{\mu}$ can be used in approximations for the long-range exchange-correlation energy $E_{xc,\erf}^{\mu}$ within ACDF approaches and $\bar{f}_{xc,\erf}^{\mu}$ can be used to construct density functional approximations for the short-range exchange-correlation energy $\bar{E}_{xc,\erf}^{\mu}$, following the same ideas as for the Coulombic functional $E_{xc,\coul}$. In comparison to $f_{xc,\coul}$, the non-locality character of $f_{xc,\erf}^{\mu}$ and $\bar{f}_{xc,\erf}^{\mu}$ tends to be attenuated by the reduction of the interaction and the modified kernels may thus be more transferable to inhomogeneous systems. In principle, these kernels can also be useful in the context of TDDFT.

The main purpose of this work is therefore the determination of the static exchange-correlation kernel of the uniform electron gas with the long-range erf interaction $f_{xc,\erf}^{\mu}$, the complement kernel $\bar{f}_{xc,\erf}^{\mu} = f_{xc,\coul} - f_{xc,\erf}^{\mu}$ being directly deducible from it. A simple approximation is constructed for $f_{xc,\erf}^{\mu}$ from its exact asymptotic behaviors which in turn require, among other things, information from the momentum distribution associated to the modified interaction. Therefore, the momentum distribution is preliminary calculated by many-body perturbation theory (MBPT), in particular using the $GW$ approximation~\cite{Hed-PR-65}. All the calculations are also performed with the Coulomb interaction which enables a check by comparison to other available data. The production of data of the same quality for $f_{xc,\coul}$ and $f_{xc,\erf}^{\mu}$ is also important to construct the complement kernel $\bar{f}_{xc,\erf}^{\mu}$.

Let's describe now the system under consideration. The uniform electron gas is described as $N$ electrons in a box of volume $\Omega$ with a uniform neutralizing background, studied in the thermodynamic limit (i.e. $N \to \infty$ and $\Omega \to \infty$ such that the density $n=N/\Omega$ remains constant). In second quantization, its Hamiltonian writes
\begin{equation}
\hat{H} = \hat{T} + \hat{V},
\end{equation}
where $\hat{T}$ is the kinetic energy operator
\begin{equation}
\hat{T} = \sum_{k\sigma} \varepsilon(k) c_{\b{k}\sigma}^\dagger c_{\b{k}\sigma},
\end{equation}
with $\varepsilon(k)=k^2/2$, and $\hat{V}$ is the electron-electron interaction
\begin{equation}
\hat{V} = \frac{1}{2\Omega} \sum_{\b{q} \ne 0} \sum_{\b{k} \sigma} \sum_{\b{k}' \sigma'} v(q) c_{\b{k}+\b{q}\sigma}^\dagger c_{\b{k}'-\b{q}\sigma'}^\dagger c_{\b{k}'\sigma'} c_{\b{k}\sigma},
\label{V}
\end{equation}
where the constant term $\b{q} = 0$ has been omitted since it cancels out with the electron-background and background-background interactions. In Eq.~(\ref{V}), $v(q)$ is the Fourier transform of an arbitrary electron-electron interaction. The standard case corresponds to the Coulomb interaction
\begin{equation}
v_{\coul}(q)=\frac{4\pi}{q^2},
\label{vcoul}
\end{equation}
while for the long-range erf interaction the Fourier transform writes
\begin{equation}
v_{\erf}^{\mu}(q)=\frac{4\pi}{q^2} e^{-q^2/(4 \mu^2)}.
\label{verf}
\end{equation}

The paper is organized as follows. In Sec.~\ref{sec:nk}, we discuss the calculation of the momentum distribution and associated properties. In Sec.~\ref{sec:fxc}, the static exchange-correlation kernel is obtained by interpolation from its limiting behaviors. In Sec.~\ref{sec:ec}, as a basic example of the use of this kernel, the long-range correlation energy of the uniform electron gas is calculated. Sec.~\ref{sec:conclusion} summarizes and concludes this work. Details of derivations and analytical parametrizations can be found in Appendices~\ref{app:virial} to~\ref{app:ABC}.

Unless otherwise stated, atomic units is assumed throughout this work.

\section{Momentum distribution}
\label{sec:nk}

One of the most reliable calculations of the momentum distribution of the Coulombic electron gas for a wide range of densities are that of Takada and Yasuhara~\cite{TakYas-PRB-91} using the effective-potential expansion method, and the quantum Monte Carlo (QMC) calculation of Ortiz and Ballone~\cite{OrtBal-PRB-94}. Here, we will consider the use of the more traditional methods of MBPT.

In MBPT, the momentum distribution $n(k)$ is expressed as
\begin{equation}
n(k)=\int_{-\infty}^{+\infty} \frac{d\w}{2\pi i} e^{i\w 0^+} G(k,\w),
\label{}
\end{equation}
where $G(k,\w)$ is the one-particle Green function, calculated in a given approximation. Daniel and Vosko~\cite{DanVos-PR-60} have calculated the momentum distribution of the Coulombic electron gas in the RPA. The fact that the RPA probably constitutes the simplest MBPT approximation for the electron gas and becomes exact in the high-density limit ($r_s \to 0$ where $r_s=(4\pi n/3)^{-1/3}$) makes it valuable for comparison to more elaborate approximations. Later, additional exchange terms have been included by Lam~\cite{Lam-PRB-71}, but the improvement over the RPA remains modest. More recently, more accurate MBPT calculations of the momentum distribution of the Coulombic electron gas have been reported using the $GW$ approximation~\cite{Hed-PR-65} at different self-consistency levels~\cite{BarHol-PRB-96,HolBar-PRB-98,Vog-THESIS-03,VogZimNee-PRB-04}. In its semi-self-consistent, $G W_0$, or fully self-consistent, $GW$, versions this method is known to be particle-conserving~\cite{BayKad-PR-61,Bay-PR-62,HolBar-PRB-98}, i.e. fulfilling the exact condition
\begin{equation}
2 \int \frac{d\b{k}}{(2\pi)^3}n(k) = n,
\label{intnk}
\end{equation}
while the non-self-consistent $G_0 W_0$ approximation is not~\cite{Sch-PRB-97}. However, even in this case, the violation of Eq.~(\ref{intnk}) is numerically small ($2\%$ at most for $r_s \le 10$ according to our own calculations). Moreover, imposition of the self-consistency does not necessary improve the momentum distribution; in particular the self-consistency increases the quasiparticle renormalization factor at the Fermi surface $Z_F$ [Eq.~(\ref{ZF})]~\cite{HolBar-PRB-98} which is suspected to be already too large at the $G_0 W_0$ level (see below). We therefore choose to compute the momentum distribution of the uniform electron gas with the modified erf interaction in the $G_0 W_0$ approximation (see Refs.~\onlinecite{BarHol-PRB-96,Vog-THESIS-03,Tou-THESIS-05} for details). For comparison purpose, calculations of the momentum distribution in the RPA (see Refs.~\onlinecite{DanVos-PR-60,Tou-THESIS-05} for details) have also been performed. The range of densities explored goes from $r_s=0.1$ to $r_s=10$ and that of interaction parameters from $\mu=0.1$ to $\mu=30$.

In order to assess the quality of the obtained momentum distributions, we look especially at two quantities that characterizes well the effect of correlations on the momentum distribution: the quasiparticle renormalization factor at the Fermi surface
\begin{equation}
Z_F = n(k\to k_F^+) - n(k\to k_F^-),
\label{ZF}
\end{equation}
where $k_F=(3\pi^2 n)^{1/3}$ is the Fermi wave-vector, and the fraction of electrons in the correlation tail (i.e., $k > k_F$)
\begin{equation}
\frac{\Delta N}{N} = \frac{2}{n} \int_{k_F}^{\infty} \frac{d k}{(2\pi)^3} 4\pi k^2 n(k).
\end{equation}

We also compute some moments of the momentum distribution: the second reduced moment $\delta_2$ related to the kinetic energy, and the fourth reduced moment $\delta_4$. The dimensionless quantity $\delta_2$ is defined as
\begin{equation}
\delta_2= \frac{\langle \hat{E}_K \rangle - \langle \hat{E}_K \rangle_0}{\langle \hat{E}_K \rangle_0},
\label{delta2}
\end{equation}
where $\hat{E}_K = (1/N) \sum_{\b{k}\sigma} \varepsilon(k) c_{\b{k}\sigma}^\dagger c_{\b{k}\sigma}$ is the kinetic energy per particle operator, $\langle \cdots \rangle_0$ and $\langle \cdots \rangle$ mean expectation values between the  non-interacting and interacting ground states, respectively. Similarly, $\delta_4$ is defined as
\begin{equation}
\delta_4= \frac{\langle \hat{E}_K^2 \rangle - \langle \hat{E}_K^2 \rangle_0}{\langle \hat{E}_K^2 \rangle_0},
\label{delta4}
\end{equation}
where $\hat{E}_K^2= (1/N) \sum_{\b{k}\sigma} \varepsilon(k)^2 c_{\b{k}\sigma}^\dagger c_{\b{k}\sigma}$. It is these moments that appear in the asymptotic expansion of the static exchange-correlation kernel (see Sec.~\ref{sec:fxc}).

We first present results for the Coulombic uniform electron gas and compare to available results in the literature in order to assess the accuracy of the $G_0 W_0$ approximation.

\begin{figure}
\includegraphics[scale=0.75]{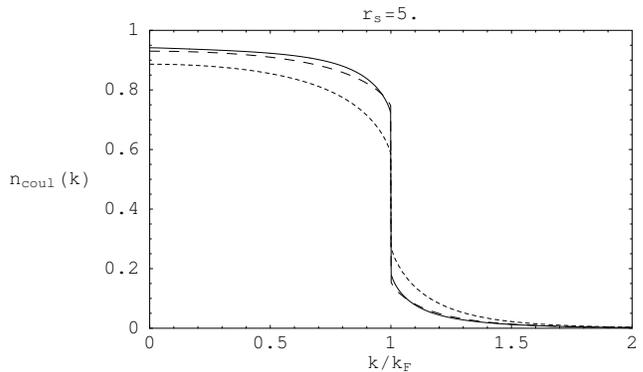}
\caption{Momentum distribution with the Coulomb interaction for $r_s=5$: GZ parametrization~\cite{GorZie-PRB-02} (solid curve), RPA (short-dashed curve) and $G_0W_0$ (long-dashed curve) calculations.
}
\label{fig:nk-coul-rs5}
\end{figure}

Fig.~\ref{fig:nk-coul-rs5} reports the momentum distribution $n_{\coul}(k)$ calculated at the RPA and $G_0 W_0$ levels for $r_s=5$. The parametrization of Gori-Giorgi and Ziesche (GZ)~\cite{GorZie-PRB-02} using accurate data from Takada and Yasuhara~\cite{TakYas-PRB-91} and known theoretical constraints is also reported. One sees that the $G_0 W_0$ calculation greatly improves the RPA and is very closed to the GZ parametrization, especially for the correlation tail.

\begin{figure}
\includegraphics[scale=0.75]{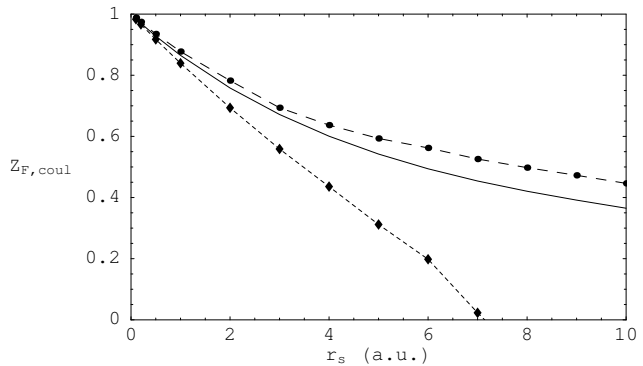}
\caption{Renormalization factor with the Coulomb interaction: GZ parametrization~\cite{GorZie-PRB-02} (solid curve), RPA (short-dashed curve) and $G_0W_0$ (long-dashed curve) calculations.
}
\label{fig:ZF-coul}
\end{figure}

The renormalization factor $Z_{F,\coul}$ is plotted in Fig.~\ref{fig:ZF-coul}. $Z_{F,\coul}$ given by the RPA is largely underestimated and even negative at very low densities ($r_s > 7$). As announced before, $G_0 W_0$ gives a renormalization factor slightly too large compared to that of the GZ parametrization which uses supposedly more accurate data of Takada and Yasuhara~\cite{TakYas-PRB-91} for $Z_{F,\coul}$.

\begin{figure}
\includegraphics[scale=0.75]{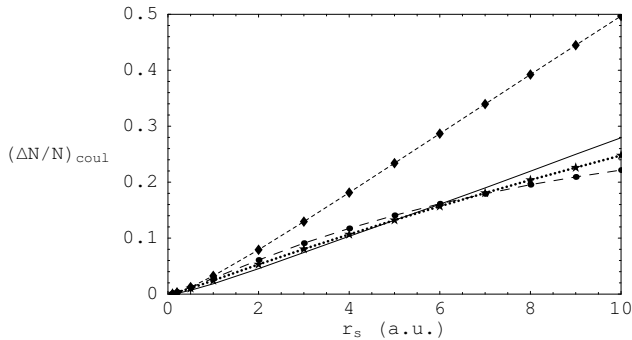}
\caption{Fraction of electrons in the correlation tail of the moment distribution with the Coulomb interaction: GZ parametrization~\cite{GorZie-PRB-02} (solid curve), RPA (short-dashed curve), $G_0W_0$ (long-dashed curve) and CCD~\cite{ReySav-IJQC-98} (dotted curve) calculations.
}
\label{fig:dNN-coul}
\end{figure}

The fraction of electrons in the correlation tail $(\Delta N/N)_{\coul}$ is reported in Fig.~\ref{fig:dNN-coul}. The RPA largely overestimates $(\Delta N/N)_{\coul}$. The $G_0 W_0$ approximation gives $(\Delta N/N)_{\coul}$ that is overall close to that given by the GZ parametrization, slightly deviating for large $r_s$ ($r_s > 8$). As an independent check, we also report in Fig.~\ref{fig:dNN-coul}, the fraction of electrons in the correlation tail extracted from a Coupled-Cluster calculation with double excitations (CCD)~\cite{ReySav-IJQC-98}. In this calculation, the energy of a uniform electron with unoccupied non-interacting levels shifted by a common gap is calculated and $(\Delta N/N)_{\coul}$ is deduced as the derivative of the energy at zero gap according to the Hellmann-Feynman theorem. The CCD calculation is close to both the GZ and $G_0 W_0$ results.

We now discuss results for the moments of the momentum distribution.

\begin{figure}
\includegraphics[scale=0.75]{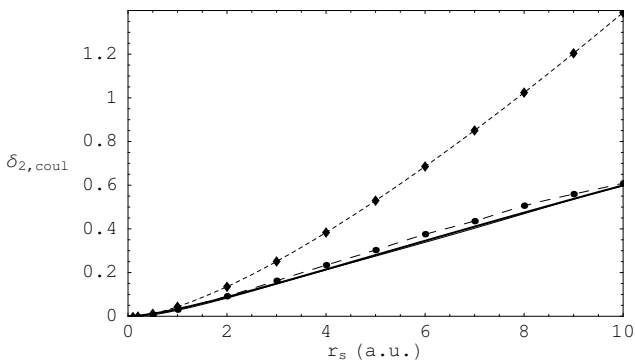}
\caption{$\delta_{2,\coul}$ with the Coulomb interaction: formula from virial theorem [Eq.~(\ref{delta2coul}), thick solid curve], GZ parametrization~\cite{GorZie-PRB-02} (solid curve superimposed with the thick solid curve), RPA (short-dashed curve) and $G_0W_0$ (long-dashed curve) calculations.
}
\label{fig:delta2-coul}
\end{figure}

Using the virial theorem, the reduced moment $\delta_{2,\coul}$ for the Coulomb interaction can be expressed exactly in term of the correlation energy of the uniform electron gas (see Appendix~\ref{app:virial}). This enables us to perform another test of the accuracy of the $G_0 W_0$ calculation. Fig.~\ref{fig:delta2-coul} compares $\delta_{2,\coul}$ obtained with the virial theorem [Eq.~(\ref{delta2coul})] using the usual Vosko-Wilk-Nusair (VWN) parametrization of the correlation energy~\cite{VosWilNus-CJP-80}, with the GZ parametrization, and with the RPA and $G_0 W_0$ calculations. The RPA calculation gives very poor results for this quantity, especially for large $r_s$. The GZ parametrization has been designed to reproduce the correct kinetic energy and consequently gives $\delta_{2,\coul}$ identical to the formula from the virial theorem. The $G_0 W_0$ approximation agrees well with this result for all $r_s$.

\begin{figure}
\includegraphics[scale=0.75]{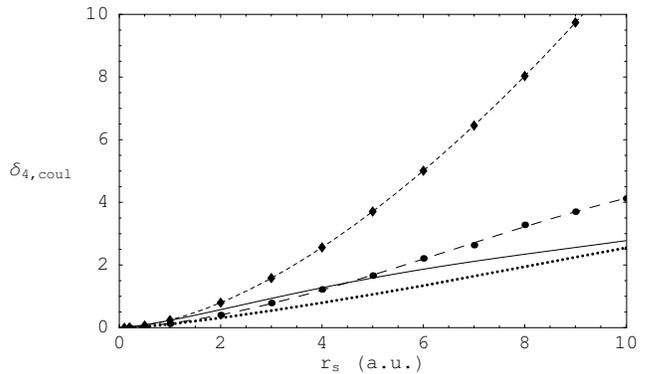}
\caption{$\delta_{4,\coul}$ with the Coulomb interaction: GZ~\cite{GorZie-PRB-02} (solid curve) and FHER~\cite{FarHeiEngRob-PRB-93} (dotted curve) parametrizations, RPA (short-dashed curve) and $G_0W_0$ (long-dashed curve) calculations.
}
\label{fig:delta4-coul}
\end{figure}

Fig.~\ref{fig:delta4-coul} shows the reduced moment $\delta_{4,\coul}$ given by the GZ parametrization and by the RPA and $G_0 W_0$ calculations. For comparison, a parametrization of $\delta_{4,\coul}$ due to Farid, Heine, Engel and Robertson (FHER)~\cite{FarHeiEngRob-PRB-93} is also reported. The $G_0 W_0$ calculation agrees with the GZ parametrization for $r_s \le 5$ but deviates from above for larger values of $r_s$. The FHER parametrization is constantly below the GZ result. These results reflect the fact that $\delta_{4,\coul}$ is difficult to compute accurately due to its important sensitivity to the correlation tail of the momentum distribution. In comparison to the RPA calculation, the $G_0 W_0$ approximation, as well as the GZ and FHER parametrizations, give ``reasonable estimates'' of $\delta_{4,\coul}$. For later use, $\delta_{4,\coul}$ given by the GZ parametrization and by the RPA and $G_0 W_0$ calculations are fitted to analytical parametrizations in Appendix~\ref{app:delta4}.

Having assessed the precision of the $G_0W_0$ approximation on the Coulombic electron gas, we now present the results for the long-range erf interaction.

\begin{figure}
\includegraphics[scale=0.75]{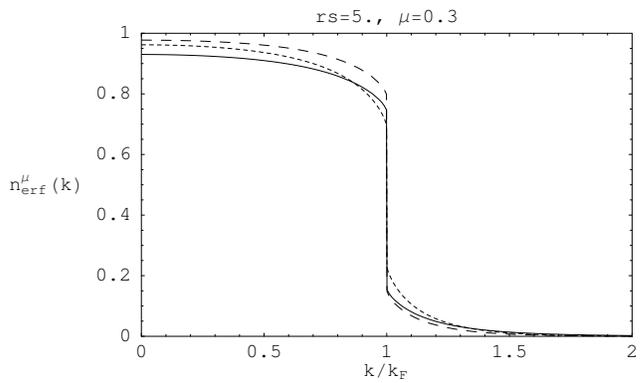}
\caption{Momentum distribution with the erf interaction for $r_s=5$ and $\mu=0.3$: RPA (short-dashed curve) and $G_0W_0$ (long-dashed curve) calculations. For comparison, the $G_0W_0$ calculation for the Coulomb case (solid curve) is also shown.
}
\label{fig:nk-erf-rs5-mu0.3}
\end{figure}

Fig.~\ref{fig:nk-erf-rs5-mu0.3} reports the momentum distribution $n^{\mu}_{\erf}$ in the RPA and $G_0W_0$ approximations for $r_s=5$ and $\mu=0.3$. For comparison with Fig.~\ref{fig:nk-coul-rs5}, the $G_0W_0$ calculation for the Coulomb interaction is also shown. The reduction of the interaction naturally brings the momentum distribution closer to the non-interacting momentum distribution $n_0(k)=\theta(k_F-k)$, even if the modifications are small at the scale of the plot, especially for the correlation tail. Also, the difference between RPA and $G_0W_0$ is reduced compared to the Coulombic case at the same density.

\begin{figure}
\includegraphics[scale=0.75]{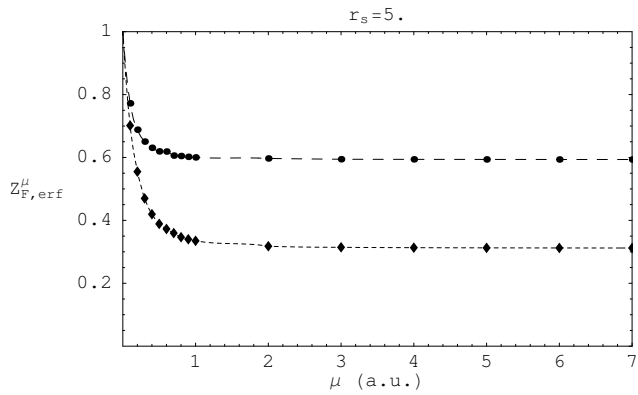}
\caption{Renormalization factor with the erf interaction for $r_s=5$: RPA (short-dashed curve) and $G_0W_0$ (long-dashed curve) calculations.
}
\label{fig:ZF-erf-rs5}
\end{figure}

The variation of the renormalization factor $Z_{F,\erf}^{\mu}$ with the interaction parameter $\mu$ is reported in Fig.~\ref{fig:ZF-erf-rs5} for the RPA and $G_0W_0$ approximation with $r_s=5$. One sees that $Z_{F,\erf}^{\mu}$ is very sensitive with respect to $\mu$ near $\mu=0$. Thus, even a very small interaction introduces significant incoherence into the evolution of an electron in the system.

\begin{figure}
\includegraphics[scale=0.75]{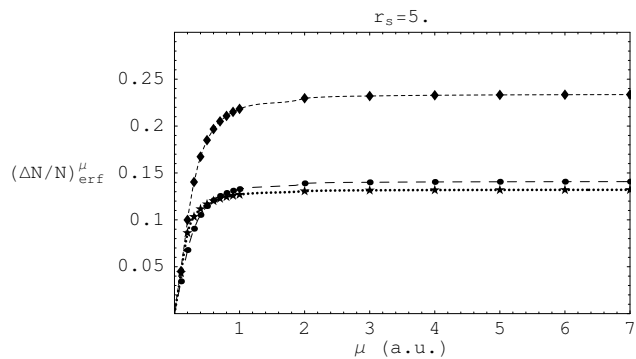}
\caption{Fraction of electrons in the correlation tail of the momentum distribution with the erf interaction: RPA (short-dashed curve), $G_0W_0$ (long-dashed curve) and CCD (dotted curve) calculations.
}
\label{fig:dNN-erf-rs5}
\end{figure}

The fraction of electrons in the correlation tail $(\Delta N/N)^{\mu}_{\erf}$ is plotted in Fig.~\ref{fig:dNN-erf-rs5} for the RPA and $G_0W_0$ calculations. A CCD calculation similar of that of Ref.~\onlinecite{ReySav-IJQC-98} but with the erf interaction is also reported. As for the Coulomb case, the $G_0W_0$ and CCD agree well in all the range of $\mu$.

The reduced moment $\delta_{2,\erf}^{\mu}$ associated to the long-range erf interaction can also be expressed in term of the correlation energy via a generalization of the virial theorem to this modified interaction (see Appendix~\ref{app:virial}). Fig.~\ref{fig:delta2-erf-rs5} compares $\delta_{2,\erf}^{\mu}$ derived from the virial theorem [Eq.~(\ref{delta2erf})] using the parametrization of the correlation energy of Ref.~\onlinecite{TouSavFla-IJQC-04} with the RPA and $G_0W_0$ results. The $G_0W_0$ calculation is in overall agreement with the formula from the virial theorem. The slight inaccuracy of $G_0W_0$ in the Coulombic limit ($\mu \to \infty$) is inherent to the method which neglects part of the short-range interactions. On the contrary, the inaccuracy for small $\mu$ is of numerical origins: when the interaction is too small, the magnitude of the correlation tail is lower than the precision of the method (which does not exceed $10^{-4}$ for the momentum distribution in the present implementation).

Finally, the reduced moment $\delta_{4,\erf}^{\mu}$ calculated in the RPA and $G_0W_0$ approximations are reported in Fig.~\ref{fig:delta4-erf-rs5} for $r_s=5$. Analytical parametrizations of these results are given in Appendix~\ref{app:delta4}. The parametrization of $\delta_{4,\erf}^{\mu}$ in the $G_0W_0$ approximation will be used in the next section for the determination of the exchange-correlation kernel.

\begin{figure}
\includegraphics[scale=0.75]{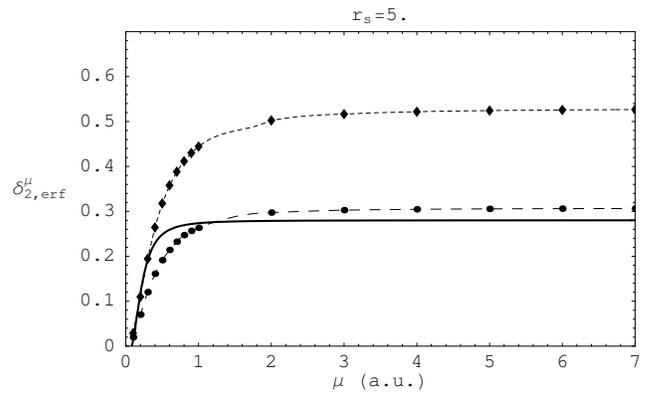}
\caption{$\delta_{2,\erf}^{\mu}$ with the erf interaction for $r_s=5$: formula from virial theorem [thick solid curve, Eq.~(\ref{delta2erf})], RPA (short-dashed curve) and $G_0W_0$ (long-dashed curve) calculations.
}
\label{fig:delta2-erf-rs5}
\end{figure}

\begin{figure}
\includegraphics[scale=0.75]{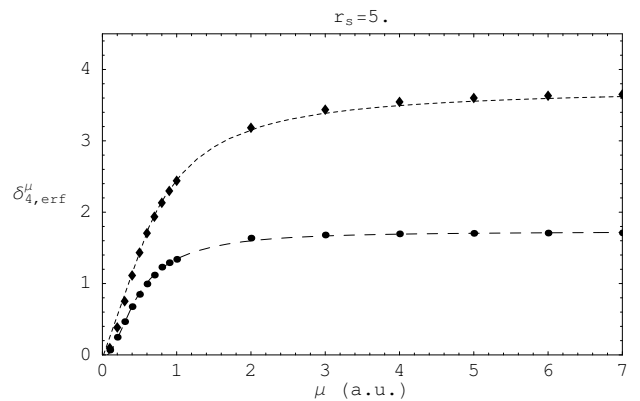}
\caption{$\delta_{4,\erf}^{\mu}$ with the erf interaction for $r_s=5$: RPA (short-dashed curve) and $G_0W_0$ (long-dashed curve) calculations.
}
\label{fig:delta4-erf-rs5}
\end{figure}

\section{Static exchange-correlation kernel}
\label{sec:fxc}

The dynamical exchange-correlation kernel $f_{xc}(k,\w)$, or equivalently the dynamical local-field factor $G(k,\w) = -v(k) f_{xc}(k,\w)$, is related to the linear dynamical (interacting) response function $\chi(k,\w)$ via the Dyson-like screening equation
\begin{equation}
\chi(k,\w) = \chi_0(k,\w) + \chi_0(k,\w)  \left[v(k) + f_{xc}(k,\w)\right] \chi(k,\w),
\label{dyson}
\end{equation}
where $\chi_0(k,\w)$ is the non-interacting dynamical response function. The static exchange-correlation kernel is $f_{xc}(k)=f_{xc}(k,\w=0)$.

The study of the local-field factor of the Coulombic electron gas has generated an abundant literature (see, e.g., Refs.~\onlinecite{Hub-PRSL-58,SinTosLanSjo-PR-68,GelVos-CJP-66,GelTay-CJP-70a,ToiWoo-PRB-70,ToiWoo-PRB-71,VasSin-PRB-72,HolAraSin-PRB-79,DevBroLem-PRB-80,BroDevLem-PRB-80,UtsIch-PRB-80a,UtsIch-PRB-80,IchUts-PRB-81,FarHeiEngRob-PRB-93,RicAsh-PRB-94,BowSugAld-PRB-94,MorCepSen-PRL-95,BreBou-PRB-96,HelGasIss-PRB-03}). The QMC simulations~\cite{BowSugAld-PRB-94,MorCepSen-PRL-95} give probably one of the most reliable static exchange-correlation kernel $f_{xc,\coul}(k)$. A remarkable feature of the QMC results is that $f_{xc,\coul}(k)$ does not have much structure, and can be essentially reproduced by combining its small-$k$ constant value for $k \lesssim 2 k_F$ and its large-$k$ two-term asymptotic expansion for $k \gtrsim 2 k_F$~\cite{MorCepSen-PRL-95}. For the case of the long-range erf interaction, accurate data are not available to check this property. However, it seems reasonable to assume that no additional structure appears in the exchange-correlation kernel when the interaction is reduced. As a first approximation, we will thus construct $f_{xc,\erf}^{\mu}(k)$ from its limiting behaviors for the case of the erf interaction too.

The limiting behaviors of $f_{xc}(k)$ for both the Coulomb and long-range erf interactions can be cast into the same form. For $k \to 0$, the static exchange-correlation kernel has the limit
\begin{equation}
f_{xc}(0) = -\frac{4\pi}{k_F^2} A,
\label{fxc0}
\end{equation}
where the dimensionless coefficient $A$ has been introduced to retain the notation of the literature~\cite{MorCepSen-PRL-95}. When $k \to \infty$, $f_{xc}(k)$ has the following asymptotic expansion (up to $k^{-2}$ order) (see Appendix~\ref{app:fxcasymptotic})
\begin{equation}
f_{xc}^{\infty}(k)  = -\frac{4\pi}{k_F^2} C - \frac{4\pi}{k^2} B,
\label{fxcinf}
\end{equation}
where $C$ and $B$ are also dimensionless quantities.

The coefficients $A$, $B$, $C$ naturally depend on the interaction chosen. $A$ is given by the compressibility sum rule (see, e.g., Refs.~\onlinecite{PinNoz-BOOK-89,Tos-INC-99}) which is valid for any electron-electron interaction. $C$ is related to the reduced second moment $\delta_2$ of the momentum distribution. The expression of $B$ involves, among other things, the second and fourth moments $\delta_2$ and $\delta_4$.

For the Coulomb interaction, $A_{\coul}$ is given by
\begin{equation}
A_{\coul} = -\frac{k_F^2}{4\pi} \frac{\partial^2 (n \varepsilon_{xc,\coul})}{\partial n^2},
\label{Acoul}
\end{equation}
with the exchange-correlation energy per particle $\varepsilon_{xc,\coul}$ taken from the usual VWN parametrization~\cite{VosWilNus-CJP-80}. The coefficients $C_{\coul}$ and $B_{\coul}$ have been calculated by Holas~\cite{Hol-INC-87,Hol-INC-91} [see Appendix~\ref{app:fxcasymptotic}, Eq.~(\ref{fxccoulkinf})] and read
\begin{eqnarray}
C_{\coul} &=& \frac{k_F^4 \delta_{2,\coul}}{5 \w_p^2},
\label{Ccoul}
\end{eqnarray}
and
\begin{eqnarray}
B_{\coul} &=& \frac{2}{3} (1-g_{\coul}(0))
+\frac{12 k_F^4 \delta_{4,\coul}}{35 \w_p^2} 
\nonumber\\
&&-\frac{4 k_F^4 (2\delta_{2,\coul}+(\delta_{2,\coul})^2)}{25\w_p^2},
\label{Bcoul}
\end{eqnarray}
where $\w_p=\sqrt{4\pi n}$ is the plasma frequency, $g_{\coul}(0)$ is the on-top pair-distribution function taken from Ref.~\onlinecite{GorPer-PRB-01}, and $\delta_{2,\coul}$ is given by the virial formula of Eq.~(\ref{delta2coul}).

\begin{figure}
\includegraphics[scale=0.75]{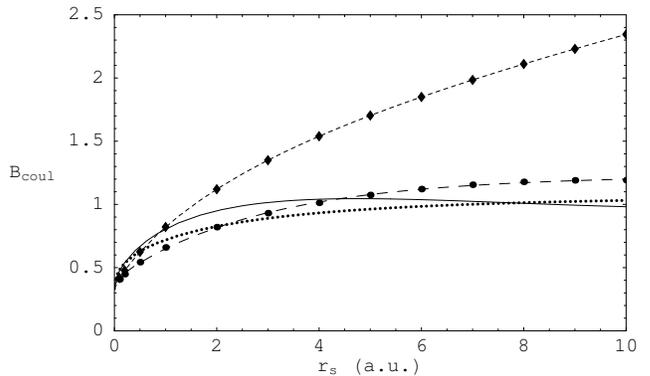}
\caption{Coefficient $B_{\coul}$ with the Coulomb interaction: GZ (solid curve) and MCS~\cite{MorCepSen-PRL-95} (dotted curve) parametrizations, RPA (short-dashed curve) and $G_0W_0$ (long-dashed curve) calculations.
}
\label{fig:B-coul}
\end{figure}

The different coefficients $B_{\coul}$ resulting from the GZ, RPA and $G_0 W_0$ parametrizations of $\delta_{4,\coul}$ [Eq.~(\ref{delta4coulfit})] are compared in Fig.~\ref{fig:B-coul}. The parametrization of Moroni, Ceperley and Senatore (MCS)~\cite{MorCepSen-PRL-95} is also reported. The $G_0 W_0$ calculation gives a reasonable estimation of $B_{\coul}$, in agreement with the GZ and MCS parametrizations.

The coefficients $A_{\coul}$, $B_{\coul}$ and $C_{\coul}$ have been calculated (using the $G_0 W_0$ parametrization of $\delta_{4,\coul}$) and fitted to simple analytical formulas for convenience in Appendix~\ref{app:ABC}.

For the long-range erf interaction, the coefficient $A_{\erf}$ is
\begin{equation}
A_{\erf}^{\mu} = -\frac{k_F^2}{4\pi} \frac{\partial^2 (n \varepsilon_{xc,\erf}^{\mu})}{\partial n^2},
\label{Aerf}
\end{equation}
where $\varepsilon_{xc,\erf}^{\mu}$ is the exchange-correlation energy per particle associated to the erf interaction taken from Ref.~\onlinecite{TouSavFla-IJQC-04}. The calculation of Holas is generalized in Appendix~\ref{app:fxcasymptotic} for the coefficients $C_{\erf}$ and $B_{\erf}$ [Eq.~(\ref{fxcerfkinf})]; they write
\begin{eqnarray}
C_{\erf}^{\mu} = \frac{k_F^4 \delta_{2,\erf}^{\mu}}{5 \w_p^2},
\label{Cerf}
\end{eqnarray}
and
\begin{eqnarray}
B_{\erf}^{\mu} &=& 
-\frac{2\mu^3}{3\pi n} \frac{\partial \varepsilon_{xc,\erf}^{\mu}}{\partial \mu} +\frac{12 k_F^4 \delta_{4,\erf}^{\mu}}{35 \w_p^2} 
\nonumber\\
&&-\frac{4 k_F^4 (2\delta_{2,\erf}^{\mu}+(\delta_{2,\erf}^{\mu})^2)}{25\w_p^2},
\label{Berf}
\end{eqnarray}
where $\delta_{2,\erf}^{\mu}$ is given by the virial formula of Eq.~(\ref{delta2erf}). The three coefficients have the correct Coulombic limit when $\mu \to \infty$: $A_{\erf}^{\mu \to \infty} = A_{\coul}$, $B_{\erf}^{\mu \to \infty} = B_{\coul}$ and $C_{\erf}^{\mu \to \infty} = C_{\coul}$. 

The coefficients $A_{\erf}^{\mu}$, $B_{\erf}^{\mu}$ and $C_{\erf}^{\mu}$ have been calculated (using the $G_0 W_0$ parametrization for $\delta_{4,\erf}^{\mu}$ [Eq.~(\ref{delta4erffit})]) and fitted to analytical formulas in Appendix~\ref{app:ABC}.

For both the Coulomb and long-range erf interactions, $f_{xc}(k)$ is then approximated over the whole range of $k$ by the simple following stepwise interpolation between the two limiting behaviors at $k \to 0$ and $k \to \infty$
\begin{eqnarray}
f_{xc}(k) = \Biggl\{
     \begin{array}{ccc}
     f_{xc}(0), && f_{xc}(0) > f_{xc}^{\infty}(k)           \\
                                                            \\
     f_{xc}^{\infty}(k), && f_{xc}(0) < f_{xc}^{\infty}(k). \\
     \end{array}
\label{fxck1}
\end{eqnarray}
The junction point between the short and long wave vector regions where $f_{xc}(0) = f_{xc}^{\infty}(k)$ is located at $k=\sqrt{\gamma} k_F$ with $\gamma = B/(A-C)$, and Eq.~(\ref{fxck1}) can be re-written in the more compact form
\begin{eqnarray}
f_{xc}(k) = -\frac{4\pi A}{k_F^2} - 4\pi \left[ \frac{B}{k^2} +\frac{C -A}{k_F^2} \right] \, \theta \left( k - \sqrt{\gamma} k_F \right).
\nonumber\\
\label{fxck}
\end{eqnarray}

\begin{figure}
\includegraphics[scale=0.75]{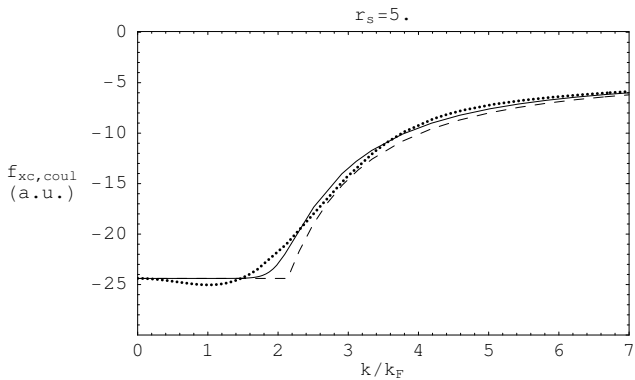}
\caption{Static exchange-correlation kernel $f_{xc,\coul}(k)$ with the Coulomb interaction for $r_s=5$: simple interpolation formula [Eq.~(\ref{fxck}), dashed curve], MCS~\cite{MorCepSen-PRL-95} (solid curve) and CDOP~\cite{CorDelOniPal-PRB-98} (dotted curve) parametrizations.
}
\label{fig:fxc-coul-rs5}
\end{figure}

In the case of the Coulomb interaction and for $r_s=5$, Fig.~\ref{fig:fxc-coul-rs5} compares interpolation of Eq.~(\ref{fxck}) with the parametrization of MCS [Eq.~(7) of Ref.~\onlinecite{MorCepSen-PRL-95} with $n=8$] and the parametrization of Corradini, Del Sole, Onida and Palummo (CDOP)~\cite{CorDelOniPal-PRB-98} based on the same QMC data. The three curves are in overall agreement, the main discrepancy arising around $k \approx 2 k_F$ where the presence of a singularity or not is still a matter of debate~\cite{FarHeiEngRob-PRB-93,RicAsh-PRB-94,MorCepSen-PRL-95,CorDelOniPal-PRB-98}. Of course, our simple model cannot be trusted for $k \approx 2 k_F$.

Taking the inverse Fourier transform of Eq.~(\ref{fxck}), we obtain the static exchange-correlation kernel for both the Coulomb and long-range erf interactions in real space
\begin{eqnarray}
f_{xc}(r) = - \frac{4\pi C}{k_F^2} \delta(\b{r}) -\frac{B}{r} \left[ 1 -\frac{2}{\pi} \Si(\sqrt{\gamma} k_F r) \right]
\nonumber\\
+ \frac{2 (C-A)}{\pi r^3 k_F^2} \left[ \sin(\sqrt{\gamma} k_F r) - \sqrt{\gamma} k_F r \cos(\sqrt{\gamma} k_F r) \right],
\label{fxcr}
\end{eqnarray}
where $\Si(x) = \int_{0}^{x} \sin(t)/t \, dt$ is the sine integral function. In Eq.~(\ref{fxcr}), the second term gives the major contribution to $f_{xc}(r)$ for $r \neq 0$, the third term being only a small correction.

\begin{figure*}
\includegraphics[scale=0.75]{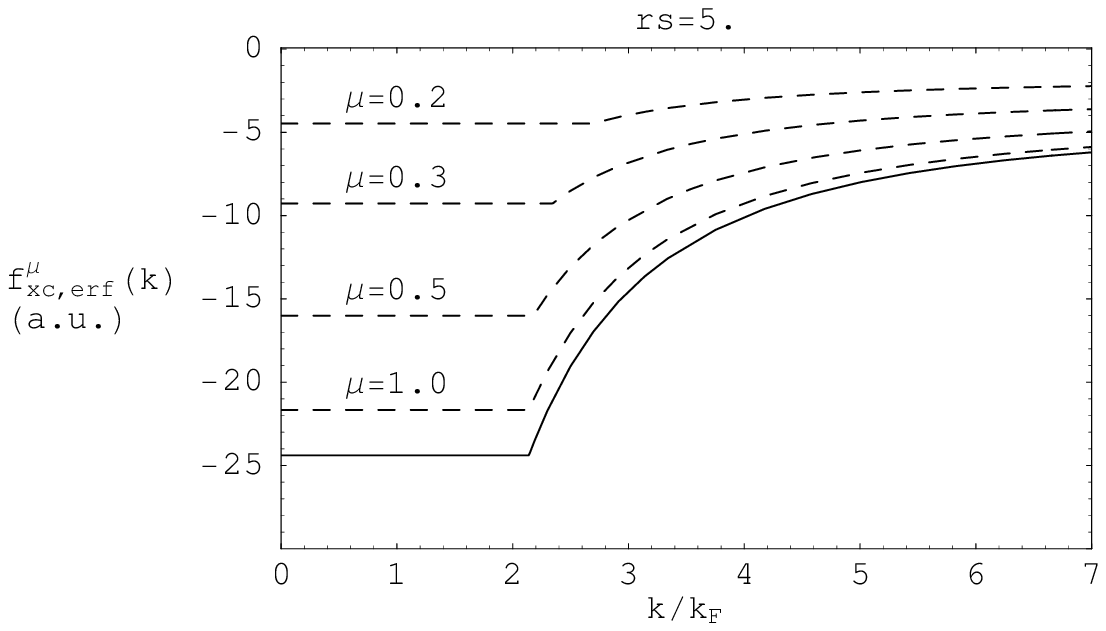}
\includegraphics[scale=0.75]{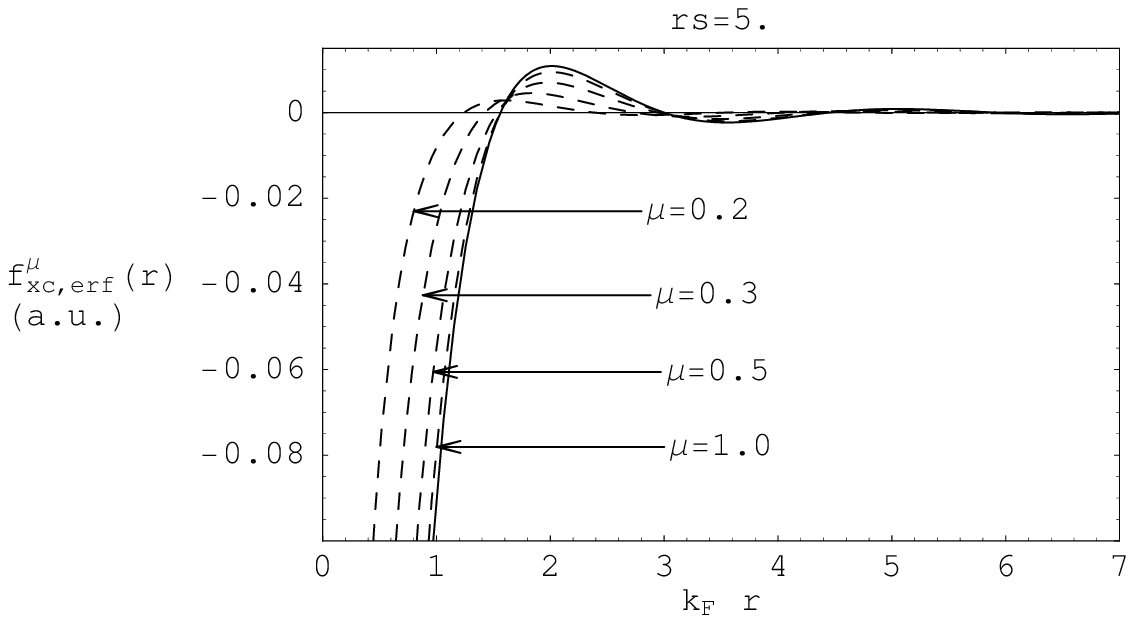}
\caption{Static exchange-correlation kernel $f_{xc,\erf}^{\mu}$ in reciprocal [Eq.~(\ref{fxck})] and direct [Eq.~(\ref{fxcr})] spaces for $r_s=5$ with the erf interaction (dashed curves) for $\mu=0.2$, $0.3$, $0.5$ and $1.0$. For comparison, the case of the Coulomb interaction (solid curve) is also shown.
}
\label{fig:fxc-erf-rs5}
\end{figure*}

Fig.~\ref{fig:fxc-erf-rs5} shows the static exchange-correlation kernel in reciprocal and real spaces for $r_s=5$ with the erf interaction for a series of interaction parameters $\mu$. $f_{xc,\erf}^{\mu}$ is naturally reduced when the interaction decreases. More precisely, in reciprocal space, $f_{xc,\erf}^{\mu}(k)$ is flatten and the junction point is slightly shifted toward larger values of $k$, while, in real space, the spatial extension of $f_{xc,\erf}^{\mu}(r)$ decreases. The kernel $f_{xc,\erf}^{\mu}$ thus becomes more and more local when $\mu$ decreases. Symmetrically, the complement kernel $\bar{f}_{xc,\erf}^{\mu} = f_{xc,\coul} - f_{xc,\erf}^{\mu}$ also becomes more and more local when $\mu$ increases (not shown).

In the limit of the Coulomb interaction, the kernel of Eq.~(\ref{fxcr}) displays Friedel-like long-range oscillations in real space of quasiperiodicity $\sqrt{\gamma} k_F \approx 2 k_F$ coming mainly from the second term in Eq.~(\ref{fxcr}). These oscillations connected with the behavior of the kernel in reciprocal space around $k \approx 2 k_F$ constitute an usual feature of models for the exchange-correlation kernel of the uniform electron gas. In our model, these oscillations are likely to be exaggerated due to the sharp cut-off in Eq.~(\ref{fxck}) and their relevance for inhomogeneous systems is questionable. However, the oscillations in our model for $f_{xc,\erf}^{\mu}(r)$ rapidly spread out and vanish when $\mu$ decreases, i.e. when the interaction is reduced, (see Fig.~\ref{fig:fxc-erf-rs5}) so that this problem disappears. Symmetrically, the oscillations in the the complement kernel $\bar{f}_{xc,\erf}^{\mu}(r)$ also disappears when $\mu$ increases. This results are consistent with the common intuition that the decrease of the interaction reduces the structure of the exchange-correlation kernel. 
Note that in the uniform electron gas the interaction can also be reduced by decreasing $r_s$ so that an equivalent way to look at $f_{xc,\erf}^{\mu}(r)$ in the limit of a small interaction, $\mu \to 0$ for a fixed $r_s$, is to look at the high-density limit, $r_s \to 0$, for a fixed $\mu$.

\section{Correlation energy from static exchange-correlation kernel}
\label{sec:ec}

As an example of the usefulness of the static exchange-correlation kernel, we follow Lein, Gross and Perdew~\cite{LeiGroPer-PRB-00} and compute the correlation energy of the uniform electron gas from it.

\begin{figure}
\includegraphics[scale=0.75]{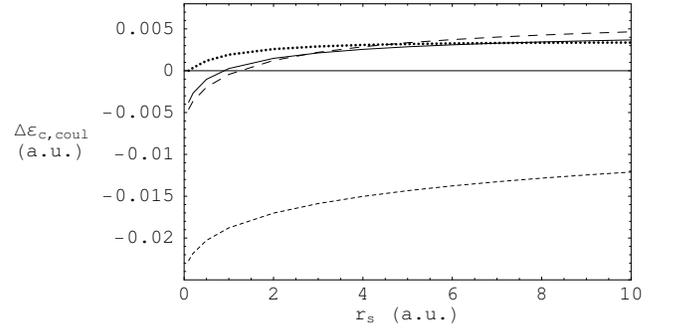}
\caption{Error in the correlation energy $\Delta\varepsilon_{c,\coul} = \varepsilon_{c,\coul}^{\text{ACFD}} - \varepsilon_{c,\coul}$ for the Coulomb interaction [Eq.~(\ref{ecACFDcoul})] with different exchange-correlation kernels: RPA [$f_{xc}=0$, short-dashed curve], MCS~\cite{MorCepSen-PRL-95} (solid curve) and CDOP~\cite{CorDelOniPal-PRB-98} (dotted curve) parametrizations and the simple interpolation formula [Eq.~(\ref{fxck}), long-dashed curve].
}
\label{fig:deltaec-coul}
\end{figure}

Indeed, the correlation energy per particle can be exactly deduced in principle from the dynamical exchange-correlation kernel using the ACFD approach. For comparison purposes, we begin by the standard case of the Coulomb interaction for which the expression is (see, e.g., Ref.~\onlinecite{LeiGroPer-PRB-00})
\begin{eqnarray}
\lefteqn{\varepsilon_{c,\coul}^{\text{ACFD}} = \frac{-1}{2\pi n} \int_{0}^{1} d\lambda \int \frac{d\b{q}}{(2\pi)^3} v_{\coul}(q)}&&
\nonumber\\
&&\times \int_{0}^{\infty} d\w \frac{\left[v^{\lambda}_{\coul}(q)+f_{xc,\coul}^{\lambda}(q,i\w) \right] \chi_0(q,i\w)^2}{1-\left[v^{\lambda}_{\coul}(q)+f_{xc,\coul}^{\lambda}(q,i\w) \right]\chi_0(q,i\w)},
\nonumber\\
\label{ecACFDcoul}
\end{eqnarray}
with the scaled interaction $v^{\lambda}_{\coul}(q) = \lambda v_{\coul}(q)$ and the associated exchange-correlation kernel $f_{xc,\coul}^{\lambda}(q,\w)$ related to $f_{xc,\coul}(q,\w)$ by a simple scaling relation~\cite{LeiGroPer-PRB-00}
\begin{equation}
f_{xc,\coul}^{\lambda}[n](q,\w) = \lambda^{-1} f_{xc,\coul}[n/\lambda^3](q/\lambda,\w/\lambda^2).
\label{fxccoullambda}
\end{equation}
We can thus use Eq.~(\ref{ecACFDcoul}) to estimate the relevance of our simple model of the static exchange-correlation kernel to obtain correlation energies, even though in theory the frequency-dependence is also needed to obtain the exact correlation energy. Fig.~\ref{fig:deltaec-coul} reports the error in the correlation obtained with Eq.~(\ref{ecACFDcoul}), $\Delta\varepsilon_{c,\coul} = \varepsilon_{c,\coul}^{\text{ACFD}} - \varepsilon_{c,\coul}$, using several approximations for $f_{xc,\coul}$: RPA ($f_{xc,\coul}=0$), MCS and CDOP parametrizations and the interpolation formula of Eq.~(\ref{fxck}). The expansion of the exact correlation energy for $r_s \to 0$ (see, e.g., Refs.~\onlinecite{GelBru-PR-57,PerWan-PRB-92,GorSacBac-PRB-00}) is $\varepsilon_{c,\coul} = C_0 \ln r_s + C_1 + \O{r_s \ln r_s}$ with $C_0=(1-\ln 2)/\pi^2$ and $C_1=-0.0469205$ while in the RPA it is $\varepsilon_{c,\coul}^{\text{RPA}} = C_0 \ln r_s + C_1^{\text{RPA}} + \O{r_s \ln r_s}$ with $C_1^{\text{RPA}}=-0.071100$. The error in the RPA correlation energy at $r_s \to 0$ is due to the difference $C_1^{\text{X}} = C_1 - C_1^{\text{RPA}}=0.024179$ corresponding to the contribution of exchange diagrams. When a static exchange-correlation kernel is used in Eq.~(\ref{ecACFDcoul}), the correlation energy is greatly improved compared to the RPA. The exchange-correlation kernel of Eq.~(\ref{fxck}) gives results very close to those given by the MCS and CDOP parametrizations, validating our simple interpolation.

We now generalize the ACFD approach to the case of the long-range erf interaction. In this case, the long-range correlation energy per particle is given by
\begin{eqnarray}
\lefteqn{\varepsilon^{\mu,\text{ACFD}}_{c,\erf} = \frac{-1}{2\pi n} \int_{0}^{1} d\lambda \int \frac{d\b{q}}{(2\pi)^3} v^{\mu}_{\erf}(q)}&&
\nonumber\\
&&\times \int_{0}^{\infty} d\w \frac{\left[v^{\mu,\lambda}_{\erf}(q)+f_{xc,\erf}^{\mu,\lambda}(q,i\w) \right] \chi_0(q,i\w)^2}{1-\left[v^{\mu,\lambda}_{\erf}(q)+f_{xc,\erf}^{\mu,\lambda}(q,i\w) \right]\chi_0(q,i\w)},
\nonumber\\
\label{ecACFDerf}
\end{eqnarray}
with the scaled interaction $v^{\mu,\lambda}_{\erf}(q) = \lambda v^{\mu}_{\erf}(q)$ and the associated exchange-correlation kernel $f_{xc,\erf}^{\mu,\lambda}(q,\w)$. The scaling relation of Eq.~(\ref{fxccoullambda}) is easily generalized to the erf interaction (see also Ref.~\onlinecite{TouSav-JJJ-XX})
\begin{equation}
f_{xc,\erf}^{\mu,\lambda}[n](q,\w) = \lambda^{-1} f_{xc,\erf}^{\mu/\lambda}[n/\lambda^3](q/\lambda,\w/\lambda^2).
\end{equation}

\begin{figure}
\includegraphics[scale=0.75]{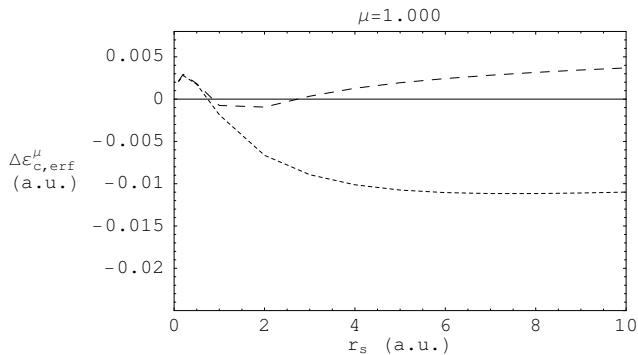}
\caption{Error in the correlation energy $\Delta\varepsilon_{c,\erf}^{\mu} = \varepsilon_{c,\erf}^{\mu,\text{ACFD}} - \varepsilon_{c,\erf}^{\mu}$ for the erf interaction [Eq.~(\ref{ecACFDerf})] with different exchange-correlation kernels: RPA [$f_{xc}=0$, short-dashed curve], the simple interpolation formula [Eq.~(\ref{fxck}), long-dashed curve], for $\mu=1$.
}
\label{fig:deltaec-erf-mu1.0}
\end{figure}

The error in the long-range correlation energy obtained with Eq.~(\ref{ecACFDerf}), $\Delta\varepsilon_{c,\erf}^{\mu} = \varepsilon_{c,\erf}^{\mu,\text{ACFD}} - \varepsilon_{c,\erf}^{\mu}$, in the RPA ($f_{xc,\erf}^{\mu}=0$) and with the exchange-correlation kernel of Eq.~(\ref{fxck}) are represented in Fig.~\ref{fig:deltaec-erf-mu1.0} for $\mu=1$. The reduction of the interaction mainly acts at small $r_s$ decreasing the error given by the RPA. As for the Coulomb case, the exchange-correlation kernel of Eq.~(\ref{fxck}) enables to correct the RPA for larger values of $r_s$.

\section{Summary and conclusion}
\label{sec:conclusion}

We have given a simple approximate expression in real and reciprocal spaces for static exchange-correlation kernel $f_{xc,\erf}^{\mu}$ of a uniform electron gas interacting with the long-range part only of the Coulomb interaction [Eq.~(\ref{fxck}) and Eq.~(\ref{fxcr}]. This expression interpolates between the exact asymptotic behaviors of the kernel at small and large wave vectors which in turn requires, among other thing, information from the momentum distribution of the uniform electron gas with the same interaction that we have calculated in the $G_0 W_0$ approximation. In the limit of the Coulomb interaction ($\mu \to \infty$), the proposed exchange-correlation kernel $f_{xc,\erf}^{\mu \to \infty} = f_{xc,\coul}$ is closed to other parametrizations of the literature based on QMC data. As a matter of course, the complement kernel $\bar{f}_{xc,\erf}^{\mu} = f_{xc,\coul} - f_{xc,\erf}^{\mu}$ associated to the short-range part of the Coulomb interaction is also obtained in this work.

In the same way that the static exchange-correlation kernel of the Coulombic uniform electron gas is an essential ingredient in ACFD and DFT approaches of the electronic correlation problem, the modified kernels $f_{xc,\erf}^{\mu}$ and $\bar{f}_{xc,\erf}^{\mu}$ can be used in alternative ACFD and DFT approaches based on a separate treatment of long-range and short-range correlations. As a simple illustration, the long-range correlation energy of the uniform electron gas has been calculated from $f_{xc,\erf}^{\mu}$. In comparison to the Coulombic kernel $f_{xc,\coul}$, the modified kernels $f_{xc,\erf}^{\mu}$ and $\bar{f}_{xc,\erf}^{\mu}$ are more local which may improve their transferability to inhomogeneous systems and therefore facilitate their use in applications. We are currently investigating such an issue.

\begin{acknowledgments}
I thank A. Savin (Paris, France) who initiated this work, P. Gori-Giorgi (Paris, France) for useful advices and in particular for Eqs.~(\ref{GPVaerf}) and~(\ref{GPVberfkinf}), F. Colonna (Paris, France), J. \'Angy\'an (Nancy, France) and J. Dobson (Griffith, Australia) for interesting discussions. I am also grateful to M. Vogt and R. Needs (Cambridge, UK) who provided me with technical details on $GW$ calculations.
\end{acknowledgments}

\appendix

\section{Kinetic energy from virial theorem}
\label{app:virial}

In this appendix, using the virial theorem, we give the expression of $\delta_2$ [Eq.~(\ref{delta2})] for the Coulomb and erf interactions.

\subsection{The virial theorem}

The virial theorem for the uniform electron gas with the Coulomb interaction writes~\cite{Mar-PR-58,Arg-PR-67}
\begin{equation}
2 t_{\coul}(r_s) + v_{\coul}(r_s) = -r_s \frac{d \varepsilon_{\coul}(r_s)}{d r_s},
\label{virialcoul}
\end{equation}
where $t_{\coul}(r_s)=t_0(r_s) + t_{c,\coul}(r_s)$ is the interacting kinetic energy per particle which can be decomposed into a non-interacting contribution $t_0(r_s)$ and a correlation contribution $t_{c,\coul}(r_s)$, $v_{\coul}(r_s)$ is the electron-electron repulsion energy per particle and $\varepsilon_{\coul}(r_s)=t_{\coul}(r_s) + v_{\coul}(r_s) = t_0(r_s) + \varepsilon_{xc,\coul}(r_s)$ is the total energy per particle which can be decomposed into the non-interacting kinetic energy $t_0(r_s)$ and the exchange-correlation energy per particle $\varepsilon_{xc,\coul}(r_s)$. Eliminating $v_{\coul}(r_s)$ for $\varepsilon_{xc,\coul}(r_s)$ in Eq.~(\ref{virialcoul}) and using the virial theorem for the non-interacting electron gas, $2 t_0(r_s) = -r_s d t_0(r_s)/d r_s$, and the virial relation for exchange only, $\varepsilon_{x,\coul}(r_s) = -r_s d \varepsilon_{x,\coul}(r_s)/d r_s$, we arrive at the virial theorem for the correlation part (see also Ref.~\onlinecite{FarHeiEngRob-PRB-93})
\begin{equation}
t_{c,\coul}(r_s) + \varepsilon_{c,\coul}(r_s)= -r_s \frac{d \varepsilon_{c,\coul}(r_s)}{d r_s}.
\label{virialccoul}
\end{equation}

For the erf interaction, the virial theorem of Eq.~(\ref{virialcoul}) is generalized as (see Ref.~\onlinecite{TouSav-JJJ-XX})
\begin{equation}
2 t^{\mu}_{\erf}(r_s) + v_{\erf}^{\mu}(r_s) -\mu \frac{\partial v_{\erf}^{\mu}(r_s)}{\partial \mu} = -r_s \frac{\partial \varepsilon^{\mu}_{\erf}(r_s)}{\partial r_s}.
\label{virialerf}
\end{equation}
Following the same steps as for the Coulomb interaction, it can be shown that the virial theorem for the correlation contributions reads now
\begin{equation}
t_{c,\erf}^{\mu}(r_s) + \varepsilon_{c,\erf}^{\mu}(r_s) -\mu \frac{\partial\varepsilon_{c,\erf}^{\mu}(r_s)}{\partial \mu} = -r_s \frac{\partial \varepsilon_{c,\erf}^{\mu}(r_s)}{\partial r_s}.
\label{virialcerf}
\end{equation}

\subsection{Expression of $\delta_2$}

Using $t_{c,\coul}(r_s)=-d (r_s \varepsilon_{c,\coul}(r_s)) /d r_s$ coming from Eq.~(\ref{virialccoul}) and $t_0(r_s)=3 k_F^2/10$, the expression of $\delta_{2,\coul}$ for the Coulomb interaction writes
\begin{eqnarray}
\delta_{2,\coul} &=& \frac{t_{c,\coul}(r_s)}{t_0(r_s)}
\nonumber\\
         &=& -\frac{10}{3k_F^2} \frac{d (r_s \varepsilon_{c,\coul}(r_s))}{d r_s}.
\label{delta2coul}
\end{eqnarray}

For the erf interaction, Eq.~(\ref{virialcerf}) gives $t_{c,\erf}^{\mu}(r_s)=-\partial (r_s \varepsilon_{c,\erf}^{\mu}(r_s)) /\partial r_s + \mu \partial \varepsilon_{c,\erf}^{\mu}(r_s) /\partial \mu$ and the expression of $\delta_{2,\erf}^{\mu}$ is
\begin{eqnarray}
\delta_{2,\erf}^{\mu} &=& \frac{t_{c,\erf}^{\mu}(r_s)}{t_0(r_s)}
\nonumber\\
         &=& \frac{10}{3k_F^2} \left( -\frac{\partial (r_s \varepsilon_{c,\erf}^{\mu}(r_s))}{\partial r_s} + \mu \frac{\partial \varepsilon_{c,\erf}^{\mu}(r_s)}{\partial \mu} \right).
\nonumber\\
\label{delta2erf}
\end{eqnarray}

\section{Analytical parametrization of $\delta_4$}
\label{app:delta4}

In this appendix, we give analytical parametrizations for the reduced moment $\delta_4$ of the momentum distribution [Eq.~(\ref{delta4})] calculated by several methods for the Coulomb and erf interactions.

\subsection{Coulomb interaction}

For the Coulomb  interaction, $\delta_{4,\coul}$ is parametrized as
\begin{equation}
\delta_{4,\coul} = \sum_{i=3}^{6} d_i x^i,
\label{delta4coulfit}
\end{equation}
where $x=\sqrt{r_s}$. A least-square fit of the result of the GZ parametrization gives $d_3=0.271191$, $d_4=-0.009998$, $d_5=-0.036383$, $d_6=0.006706$. For the RPA calculation we obtain $d_3=0.093623$, $d_4=0.194288$, $d_5=0.051445$, $d_6=0.005449$, and for the $G_0 W_0$ calculation, $d_3=0.126362$, $d_4=0.001428$, $d_5=0.014278$, $d_6=-0.004522$.

\subsection{erf interaction}
For the erf interaction, we take the parametrization
\begin{equation}
\delta_{4,\erf}^{\mu} = \sum_{i=3}^{6} e_i x^i,
\label{delta4erffit}
\end{equation}
with $e_i = (e_{i1} \mu + e_{i2} \mu^2)/(1+e_{i3} \mu^2)$. The optimal parameters reproducing the RPA calculation are $e_{31}=-0.690727$, $e_{32}=0.393525$, $e_{33}=3.398631$, $e_{41}=0.231529$, $e_{42}=0.885698$, $e_{43}=5.999882$, $e_{51}=0.001233$, $e_{52}=-0.061940$, $e_{53}=2.887302$, $e_{61}=0.001083$, $e_{62}=-0.000060$, $e_{63}=0.391712$. Those for the $G_0 W_0$ calculation are $e_{31}=-0.525472$, $e_{32}=0.417720$, $e_{33}=5.102281$, $e_{41}=0.223039$, $e_{42}=0.316492$, $e_{43}=5.984490$, $e_{51}=0.002486$, $e_{52}=-0.000770$, $e_{53}=0.265086$, $e_{61}=-0.004998$, $e_{62}=-0.003175$, $e_{63}=1.191797$.

\section{Asymptotic expansion of the static exchange-correlation kernel}
\label{app:fxcasymptotic}

In this appendix, following the procedure proposed by Holas~\cite{Hol-INC-87,Hol-INC-91}, we derive the asymptotic expansion of the static exchange-correlation kernel $f_{xc}(k)$ of the uniform electron gas for $k \to \infty$ for the erf interaction. For comparison, results for the Coulomb interaction are also given. Contrary to the rest of this work and except for Eqs.~(\ref{fxccoulkinf}) and~(\ref{fxcerfkinf}), momentum are expressed in units of $k_F$, energies and frequencies in units of $k_F^2$ throughout this appendix.

\subsection{Frequency moments of the linear response function}

Knowledge of the frequency moments of the frequency-dependent linear response function $\chi(k,\w)$ is useful to study the asymptotic behavior of the static response function for $k \to \infty$. These moments are defined as
\begin{equation}
M_{l}(k) = \frac{2}{\pi} \int_{0}^{\infty} \Im \chi(k,\w) \w^l d\w.
\end{equation}
The zeroth, first and third moments are well-known~\cite{PatVas-PRB-72,Hol-INC-87,Hol-INC-91}
\begin{equation}
M_{0}(k) = - 2 n S(k),
\label{M0}
\end{equation}
\begin{equation}
M_{1}(k) = -  n k^2,
\label{M1}
\end{equation}
\begin{eqnarray}
M_{3}(k) &=& - n k^2 \biggl( \frac{k^4}{2} +2 k^2 \langle \hat{E}_K \rangle + n k^2 v(k) 
\nonumber\\
&&-4\pi n G^{\text{PV}}(k) \biggl),
\label{M3}
\end{eqnarray}
where $S(k)$ is the static structure factor, $\hat{E}_K = (1/N) \sum_{\b{k}\sigma} \varepsilon(k) c_{\b{k}\sigma}^\dagger c_{\b{k}\sigma}$ is the kinetic energy per particle operator, $\langle \cdots \rangle$ means expectation value between the interacting ground state, and $G^{\text{PV}}(k)$ is the ``Pathak-Vashishta''~\cite{PatVas-PRB-72} (or ``Niklasson''~\cite{Nik-PRB-74}) local-field factor
\begin{equation}
G^{\text{PV}}(k) = G^{\text{PV}}_{a}(k) + G^{\text{PV}}_{b}(k),
\end{equation}
with
\begin{equation}
G^{\text{PV}}_{a}(k) = \frac{1}{n} \int \frac{d\b{p}}{(2\pi)^3} \frac{(\b{k} \cdot \b{p})^2}{k^2} \frac{v(p)}{4\pi} \left( S(p) - 1 \right),
\end{equation}
and
\begin{equation}
G^{\text{PV}}_{b}(k) = -\frac{1}{n} \int \frac{d\b{p}}{(2\pi)^3} \frac{(\b{k} \cdot \b{p})^2}{k^2} \frac{v(p)}{4\pi} \left( S(|\b{k}-\b{p}|) -1 \right).
\end{equation}
The asymptotic expansion of $M_{0}(k)$ as $k \to \infty$ is given by that of $S(k)$~\cite{Kim-PRA-73}
\begin{equation}
M_{0}(k) = - 2 n \left(1+\frac{C}{k^4}+o\left(\frac{1}{k^4}\right)\right).
\label{M0kinf}
\end{equation}
For the Coulomb interaction, $C$ is directly related to the cusp condition: $C_{\coul}=-2 \w_p^2 g'(0)=-2 \w_p^2 g(0)$ where $\w_p=\sqrt{4\pi n}$ is the plasma frequency,  $g(0)$ and $g'(0)$ are the on-top values of the pair-distribution function and of its spherical-average derivative. For the cusp-less erf interaction, $C$ vanishes: $C_{\erf}=0$.

The asymptotic expansion of $M_{3}(k)$ writes
\begin{eqnarray}
M_{3}(k) &=& -  \frac{n k^6}{4} \biggl( 1 + \frac{8}{k^2} \langle \hat{E}_K \rangle + \frac{4}{k^4} n v(k)
\nonumber\\
&& -\frac{16}{k^4} \pi n G^{\text{PV}}(k \to \infty) + o\left( \frac{1}{k^4} \right) \biggl),
\label{M3kinf}
\end{eqnarray}
where the limit $G^{\text{PV}}(k \to \infty)$ is determined as follows.

$G^{\text{PV}}_{a}(k)$ is actually independent of $k$ and writes
\begin{equation}
G^{\text{PV}}_{a} = \frac{1}{3} \frac{1}{n} \int \frac{d\b{p}}{(2\pi)^3} \frac{p^2 v(p)}{4\pi} \left( S(p) - 1 \right),
\end{equation}
which gives, for the Coulomb interaction,
\begin{eqnarray}
G^{\text{PV}}_{a,\coul} &=& \frac{1}{3} \frac{1}{n} \int \frac{d\b{p}}{(2\pi)^3} \left( S_{\coul}(p) - 1 \right)
\nonumber\\
&=& \frac{1}{3} \left( g_{\coul}(0) - 1 \right),
\end{eqnarray}
and, for the erf interaction,
\begin{eqnarray}
G^{\text{PV},\mu}_{a,\erf} &=& \frac{1}{3} \frac{1}{n} \int \frac{d\b{p}}{(2\pi)^3} e^{-p^2/(4\mu^2)} \left( S_{\erf}^{\mu}(p) - 1 \right)
\nonumber\\
&=& \frac{\mu^3}{6\pi n} \int \frac{d\b{p}}{(2\pi)^3} \frac{\partial v_{\erf}^{\mu}(p)}{\partial \mu} \left( S_{\erf}^{\mu}(p) - 1 \right)
\nonumber\\
&=& \frac{\mu^3}{3\pi n} \frac{\partial \varepsilon_{xc,\erf}^{\mu}}{\partial \mu},
\label{GPVaerf}
\end{eqnarray}
where the Hellmann-Feynman theorem has been used for the last line, and $\varepsilon_{xc,\erf}^{\mu}$ is the exchange-correlation energy per particle with the erf interaction.

After a trivial variable transformation, the limit of $G^{\text{PV}}_{b}$ as $k \to \infty$ is easily seen to be
\begin{eqnarray}
G^{\text{PV}}_{b}(k \to \infty) &=& -\frac{1}{n} \int \frac{d\b{p}}{(2\pi)^3} \frac{(\b{k}-\b{p})^2 v(|\b{k}-\b{p}|)}{4\pi} 
\nonumber\\
&&\times \left( S(p) - 1 \right),
\nonumber\\
\end{eqnarray}
which gives, for the Coulomb interaction,
\begin{eqnarray}
G^{\text{PV}}_{b,\coul}(k \to \infty) &=& -\frac{1}{n} \int \frac{d\b{p}}{(2\pi)^3} \left( S_{\coul}(p) - 1 \right)
\nonumber\\
&=& - \left( g_{\coul}(0) -1 \right),
\end{eqnarray}
and, for the erf interaction,
\begin{eqnarray}
G^{\text{PV},\mu}_{b,\erf}(k \to \infty) &=& - \frac{1}{n} \int \frac{d\b{p}}{(2\pi)^3} e^{-(\b{k}-\b{p})^2/(4\mu^2)} 
\nonumber\\
&&\times \left( S_{\erf}^{\mu}(p) - 1 \right)
\nonumber\\
&=& - e^{-k^2/(4\mu^2)} \left( g_{\erf}^{\mu}(0) -1 \right),
\label{GPVberfkinf}
\end{eqnarray}
where $g_{\erf}^{\mu}(0)$ is the on-top pair-distribution function associated to the erf interaction.

\subsection{Static linear response function}

The static linear response function $\chi(k,0)$ can be expressed by the spectral representation
\begin{equation}
\chi(k,0) = \frac{2}{\pi} \int_{0}^{\infty} \frac{\Im \chi(k,\w)}{\w} d\w,
\end{equation}
i.e., we simply have $\chi(k,0) = M_{-1}(k)$. Using this result, Holas~\cite{Hol-INC-87,Hol-INC-91} showed that the asymptotic expansions for large $k$ of $M_{0}(k)$, $M_{1}(k)$ and $M_{3}(k)$ [Eqs.~(\ref{M0kinf}),~(\ref{M1}) and~(\ref{M3kinf})] are sufficient to determine the asymptotic expansion of $\chi(k,0)$ up to $k^{-4}$ order
\begin{eqnarray}
\chi(k,0) &=& \frac{-4n}{k^2} \biggl[ 1 + \frac{8}{3} \langle \hat{E}_K \rangle \frac{1}{k^2} + o\left(\frac{1}{k^2}\right)\biggl].
\label{chik4inf}
\end{eqnarray}
He then determined the following terms in the asymptotic expansion of $\chi(k,0)$ by (not rigorously) inferring them as follows. The following term in $k^{-3}$ in the square-bracket of Eq.~(\ref{chik4inf}) is set to zero by a simple argument involving the continuity and the likely non-oscillating behavior of the moments $M_{l}(k)$ with respect to $l$. The next term in $k^{-4}$ is inferred from the asymptotic expansion of the static response function calculated to first-order with respect to the electron-electron interaction. We follow the same procedure.

The (zeroth-order) Lindhard static response function $\chi_0(k,0)$ is well-known. Its asymptotic expansion for large $k$ writes
\begin{eqnarray}
\chi_0(k,0) &=& \frac{-4n}{k^2} \biggl[ 1 + \frac{8}{3} \langle \hat{E}_K \rangle_0 \frac{1}{k^2} 
\nonumber\\
&&+\frac{64}{5} \langle \hat{E}_K^2 \rangle_0 \frac{1}{k^4} + o\left(\frac{1}{k^4}\right)\biggl],
\label{chi0kinf}
\end{eqnarray}
where $\hat{E}_K^2= (1/N) \sum_{\b{k}\sigma} \varepsilon(k)^2 c_{\b{k}\sigma}^\dagger c_{\b{k}\sigma}$ and $\langle \cdots \rangle_0$ means expectation value between the non-interacting ground state. 

The asymptotic expansion of the first-order correction to the static response function $\chi_1(k,0)$ has been calculated for the Coulomb interaction by Geldart and Taylor~\cite{GelTay-CJP-70,GelTay-CJP-70a} (see also Ref.~\onlinecite{HolAraSin-PRB-79}). This result is conveniently generalized to an arbitrary interaction as
\begin{eqnarray}
\chi_1(k,0) &=& \frac{-4n}{k^2} \biggl[ -4\w_p^2  \left( 2 G^{\text{PV}}_{a,\text{HF}} +\frac{k^2v(k)}{4\pi} \right)  \frac{1}{k^4} 
\nonumber\\
&&+ o\left(\frac{1}{k^4}\right)\biggl],
\end{eqnarray}
with
\begin{equation}
G^{\text{PV}}_{a,\text{HF}} = \frac{1}{3} \frac{1}{n} \int \frac{d\b{p}}{(2\pi)^3} \frac{p^2 v(p)}{4\pi} \left( S_{\text{HF}}(p) - 1 \right),
\end{equation}
and the Hartree-Fock (HF) static structure factor
\begin{equation}
S_{\text{HF}}(p) = 1 - \frac{2}{n} \int \frac{d\b{k}}{(2\pi)^3} n_0(k) n_0(|\b{k}+\b{p}|).
\end{equation}
For the Coulomb interaction, we simply have $G^{\text{PV}}_{a,\text{HF},\coul} = ( g_{\text{HF},\coul}(0) - 1 )/3$ where $g_{\text{HF},\coul}(0)=1/2$ is the HF on-top pair-distribution function. For the erf interaction, $G^{\text{PV},\mu}_{a,\text{HF},\erf} = \mu^3/(3\pi n) \partial \varepsilon_{x,\erf}^{\mu}/\partial \mu$ where $\varepsilon_{x,\erf}^{\mu}$ is the exchange energy per particle associated to this modified interaction.

Following Holas~\cite{Hol-INC-87,Hol-INC-91}, we then infer the following term in $k^{-4}$ in the square-bracket of Eq.~(\ref{chik4inf}) from the corresponding term in the expansion of $\chi_0(k,0)+\chi_1(k,0)$ with the substitutions $\langle \hat{E}_K^2 \rangle_0 \to \langle \hat{E}_K^2 \rangle$ and $G^{\text{PV}}_{a,\text{HF}} \to G^{\text{PV}}_{a}$. We thus finally obtain
\begin{eqnarray}
\chi(k,0) &=& \frac{-4n}{k^2} \biggl[ 1 + \frac{8}{3} \langle \hat{E}_K \rangle \frac{1}{k^2} + \biggl\{ \frac{64}{5} \langle \hat{E}_K^2 \rangle
\nonumber\\
&& -4\w_p^2  \left( 2 G^{\text{PV}}_{a} +\frac{k^2v(k)}{4\pi} \right)  \biggl\} \frac{1}{k^4}
+ o\left(\frac{1}{k^4}\right)\biggl].
\nonumber\\
\label{chikinf}
\end{eqnarray}
Observe that the form of the asymptotic expansion of $\chi(k,0)$ [Eq.~(\ref{chikinf})] does depend explicitly on the interaction $v(k)$ and is therefore different for the Coulomb and erf interactions.

\subsection{Static exchange-correlation kernel}

The static exchange-correlation kernel $f_{xc}(k)$ writes [see Eq.~(\ref{dyson})]
\begin{equation}
f_{xc}(k) = \chi_0(k,0)^{-1} - \chi(k,0)^{-1} - v(k),
\label{fxckdef}
\end{equation}
and its asymptotic expansions for large $k$ is therefore determined from the expansions of $\chi_0(k,0)$ and $\chi(k,0)$ [Eqs.~(\ref{chi0kinf}) and~(\ref{chikinf})]. Introducing the quantities $\delta_2=(\langle \hat{E}_K \rangle - \langle \hat{E}_K \rangle_0)/\langle \hat{E}_K \rangle_0$ and $\delta_4=(\langle \hat{E}_K^2 \rangle - \langle \hat{E}_K^2 \rangle_0)/\langle \hat{E}_K^2 \rangle_0$ with $\langle \hat{E}_K \rangle_0=3/10$ and $\langle \hat{E}_K^2 \rangle_0=3/28$, we obtain (with $k$ in units of $k_F$ and $\w_p$ in units of $k_F^2$)
\begin{eqnarray}
f_{xc}(k) &=& \frac{-4\pi\delta_2}{5 \w_p^2} -\frac{4\pi}{k^2} \left( - 2 G^{\text{PV}}_{a} +\frac{12 \delta_4}{35 \w_p^2}
-\frac{4(2\delta_2+\delta_2^2)}{25\w_p^2} \right)
\nonumber\\
&& + o\left(\frac{1}{k^2} \right).
\nonumber\\
\label{fxckinf}
\end{eqnarray}
The term explicitly depending on $v(k)$ in the asymptotic expansion of $\chi(k,0)$ [Eq.~(\ref{chikinf})] exactly cancels with the Hartree kernel $v(k)$ in the expression of $f_{xc}(k)$ [Eq.~(\ref{fxckdef})]. Therefore, the asymptotic expansion of $f_{xc}(k)$ [Eq.~(\ref{fxckinf})] do has the same form for the Coulomb and erf interactions. Note however that the local field factor $G(k)=-v(k)f_{xc}(k)$ involves explicitly the interaction and has therefore different asymptotic behaviors for the Coulomb and erf interactions.

Coming back to atomic units now and specializing to the Coulomb interaction, Eq.~(\ref{fxckinf}) becomes
\begin{eqnarray}
\lefteqn{f_{xc,\coul}(k) = \frac{-4\pi k_F^2 \delta_{2,\coul}}{5 \w_p^2} -\frac{4\pi}{k^2} \biggl( \frac{2}{3} (1-g_{\coul}(0))}&&
\nonumber\\
&&+\frac{12 k_F^4 \delta_{4,\coul}}{35 \w_p^2} -\frac{4 k_F^4 (2\delta_{2,\coul}+(\delta_{2,\coul})^2)}{25\w_p^2} \biggl) + o\left(\frac{1}{k^2} \right).
\nonumber\\
\label{fxccoulkinf}
\end{eqnarray}
while for the erf interaction, we have
\begin{eqnarray}
\lefteqn{f_{xc,\erf}^{\mu}(k) = \frac{-4\pi k_F^2 \delta_{2,\erf}^{\mu}}{5 \w_p^2} -\frac{4\pi}{k^2} \biggl( -\frac{2\mu^3}{3\pi n} \frac{\partial \varepsilon_{xc,\erf}^{\mu}}{\partial \mu}}&&
\nonumber\\
&& +\frac{12 k_F^4 \delta_{4,\erf}^{\mu}}{35 \w_p^2} -\frac{4 k_F^4 (2\delta_{2,\erf}^{\mu}+(\delta_{2,\erf}^{\mu})^2)}{25\w_p^2} \biggl) 
+ o\left(\frac{1}{k^2} \right).
\nonumber\\
\label{fxcerfkinf}
\end{eqnarray}

\section{Analytical parametrizations of $A$, $B$ and $C$}
\label{app:ABC}

To facilitate the evaluation of the static exchange-correlation kernel for the Coulomb and erf interaction, we give analytical parametrizations of the coefficients $A$, $B$ and $C$ determining its limiting behaviors for small and large $k$ (see Sec.~\ref{sec:fxc}).

\subsection{Coulomb interaction}

For the Coulomb interaction, the coefficient $A$ [Eq.~(\ref{Acoul})] is parametrized as follows
\begin{equation}
A_{\coul} = \sum_{i=0}^{6} a_i x^i,
\label{Acoulfit}
\end{equation}
with $x=\sqrt{r_s}$ and the fitted parameters $a_0=0.250019$, $a_1=-0.000162$, $a_2=0.013441$, $a_3=-0.003591$, $a_4=0.000380$, $a_5=0.000002$, $a_6=-0.000003$.

For the coefficient $B$ (Eq.~\ref{Bcoul}), we adopt the analytical form of Moroni, Ceperley and Senatore~\cite{MorCepSen-PRL-95} incorporating the correct limit $B(r_s \to 0)=1/3$
\begin{equation}
B_{\coul} = \frac{1+b_1 x +b_2 x^3}{3+b_3 x + b_4 x^3},
\label{Bcoulfit}
\end{equation}
but the parameters are refitted with the parametrization of $\delta_4$ [Eq.~(\ref{delta4coulfit})] with the $G_0 W_0$ data: $b_1=0.721543$, $b_2=0.317320$, $b_3=-0.133379$, $b_4=0.269494$.

The coefficient $C$ [Eq.~(\ref{Ccoul})] is parametrized as
\begin{equation}
C_{\coul} = \frac{\sum_{i=1}^{6} c_i x^i}{1+\sum_{i=1}^{4} f_i x^i},
\label{Ccoulfit}
\end{equation}
with $c_1=0.002127$, $c_2=0.169597$, $c_3=0.450771$, $c_4=-0.023265$, $c_5=0.001855$, $c_6=-0.000069$, $f_1=7.062604$, $f_2=8.589773$, $f_3=2.747407$, $f_4=0.648920$.

\subsection{erf interaction}

For the erf interaction, $A_{\erf}^{\mu}$ [Eq.~(\ref{Aerf})] is represented by the analytical parametrization
\begin{equation}
A_{\erf}^{\mu} = \frac{\sum_{i=2}^{6} g_i x^i}{1+h_4 x^4},
\label{Aerffit}
\end{equation}
with $g_i=g_{i1} \mu + g_{i2} \mu^2$ and $h_4=g_{41} \mu + h_{42} \mu^2$. The fitted parameters are $g_{21}=-0.029315$, $g_{22}=-0.000927$, $g_{31}=0.061867$, $g_{32}=0.010970$, $g_{41}=-0.053761$, $g_{42}=0.078580$, $g_{51}=0.012970$, $g_{52}=0.014669$, $g_{61}=-0.001232$, $g_{62}=-0.000891$, $h_{41}=-0.025963$, $h_{42}= 0.389673$.

The coefficient $B_{\erf}^{\mu}$ [Eq.~(\ref{Berf})] is parametrized as
\begin{equation}
B_{\erf}^{\mu} = \frac{\sum_{i=0}^{3} j_i x^i}{1+k_1 x + k_3 x^3},
\label{Berffit}
\end{equation}
where $j_i = (j_{i1} \mu + j_{i2} \mu^2)/(1+j_{i3} \mu^2)$ and $k_i = (k_{i1} \mu + k_{i2} \mu^2)/(1+k_{i3} \mu^2)$. Using the parametrization of Eq.~(\ref{delta4erffit}) for $\delta_4^{\mu}$ in the $G_0 W_0$ approximation, the fitted parameters obtained are $j_{01}=0.010533$, $j_{02}=-0.002640$, $j_{03}=0.314403$, $j_{11}=-0.143455$, $j_{12}=0.046302$, $j_{13}=0.014315$, $j_{21}=-0.415043$, $j_{22}=0.194149$, $j_{23}=0.078849$, $j_{31}=0.164085$, $j_{32}=0.925083$, $j_{33}=0.491612$, $k_{11}=0.291633$, $k_{12}=0.102905$, $k_{13}=0.013551$, $k_{31}=0.014324$, $k_{32}=0.714449$, $k_{33}=0.473130$.

Finally, the coefficient $C_{\erf}^{\mu}$ [Eq.~(\ref{Cerf})] is parametrized as
\begin{equation}
C_{\erf}^{\mu} = \frac{\sum_{i=1}^{6} l_i x^i}{1+m_4 x^4},
\label{Cerffit}
\end{equation}
where $l_i = (l_{i1} \mu + l_{i2} \mu^2)/(1+l_{i3} \mu^2)$ and $m_4 = (m_{41} \mu + m_{42} \mu^2)/(1+m_{43} \mu^2)$. The fitted parameters are $l_{11}=-0.039662$, $l_{12}=0.002346$, $l_{13}=0.405492$, $l_{21}=0.187782$, $l_{22}=0.066673$, $l_{23}=0.759503$, $l_{31}=-0.270823$, $l_{32}=-0.083482$, $l_{33}=0.511806$, $l_{41}=0.133824$, $l_{42}=0.061658$, $l_{43}=0.348085$, $l_{51}=-0.027392$, $l_{52}=0.024384$, $l_{53}=0.261739$, $l_{61}=0.001877$, $l_{62}=-0.005601$, $l_{63}=0.351272$, $m_{41}=-0.090916$, $m_{42}=1.238974$, $m_{43}=0.251286$.

\bibliographystyle{apsrev}
\bibliography{biblio}

\end{document}